\documentclass[11pt,a4paper]{article}
\pdfoutput=1
\usepackage{jheppub}
\usepackage{gensymb}
\usepackage{subfigure}
\usepackage{amssymb,amsmath}
\usepackage{graphicx}
\usepackage{color}
\usepackage{cancel}
\usepackage[colorlinks=true
,urlcolor=blue
,citecolor=blue
,linkcolor=blue
,pagecolor=blue
,linktocpage=true
,pdfproducer=medialab
]{hyperref}
\usepackage[section]{placeins}
 \usepackage[numbers]{natbib}
\usepackage{notoccite}
\makeatletter \renewcommand{\@dotsep}{10000} \makeatother
\def\be{\begin{equation}}
\def\ee{\end{equation}}
\def\bea{\begin{eqnarray}}
\def\eea{\end{eqnarray}}
\def\bi{\begin{itemize}}
\def\ei{\end{itemize}}

\usepackage[nodisplayskipstretch]{setspace}

\newcommand{\beq}{\begin{equation}}
\newcommand{\eeq}{\end{equation}}

\DeclareUnicodeCharacter{2212}{-}
\DeclareUnicodeCharacter{2217}{-}
\begin{document}

\begin{titlepage}
\pagestyle{empty}

\vspace*{0.2in}
\begin{center}
{\Large \bf    The 28 GeV Dimuon Excess in Lepton Specific THDM}\\
\vspace{1cm}
{Ali  \c{C}i\c{c}i$^{a,}$\footnote{E-mail: 501507007@ogr.uludag.edu.tr},
Shaaban Khalil$^{b,}$\footnote{E-mail: skhalil@zewailcity.edu.eg},
B\"{u}\c{s}ra Ni\c{s}$^{a,}$\footnote{Email: 501507008@ogr.uludag.edu.tr} 
 and
Cem Salih $\ddot{\rm U}$n$^{a,c,}$\footnote{E-mail: cemsalihun@uludag.edu.tr}}
\vspace{0.5cm}

{\it
$^a$Department of Physics, Bursa Uluda\~{g} University, TR16059 Bursa, Turkey \\
$^b$ Center of Fundamental Physics, Zewail City of Science and Technology, Sheikh Zayed, 12588 Giza, Egypt \\
$^c$ Departamento de Ciencias Integradas y Centro de Estudios Avanzados en F\'{i}sica Matem\'aticas y Computación, Campus del Carmen, Universidad de Huelva, Huelva 21071, Spain}

\end{center}

\vspace{0.5cm}
\begin{abstract}

We explore the Higgs mass spectrum in a class of Two Higgs Doublet Models (THDMs) in which a scalar $SU(2)_{L}$ doublet interacts only with quarks, while the second one interacts only with leptons. The spectrum includes two CP-even Higgs bosons, either of which can account for the SM-like Higgs boson, and the spectra involving light Higgs bosons receive strong impacts from the LEP results and the current collider analyses. We find that a consistent spectrum can involve a CP-odd Higgs boson as light as about 10 GeV, while the lightest CP-even Higgs boson cannot be lighter than about 55 GeV when $m_{A}\sim 28$ GeV.  These analyses can rather bound the low $\tan\beta$ region which can also accommodate an observed excess in dimuon events at $m_{\mu\mu}\sim 28$ GeV. A lepton-specific class of THDMs (LS-THDM) can predict such an excess through $A\rightarrow \mu\mu$ decays, while the solutions can be constrained by the $A\rightarrow \tau\tau$ mode. After constraining the solutions with the consistent ranges of $\sigma(pp\rightarrow bbA \rightarrow bb\tau\tau)$, a largest excess at about $1.5\sigma$ at 8 TeV center of mass (COM) energy and $2\sigma$ at 13 TeV COM is observed for $\tan\beta \sim 12$ and $m_{A}\sim 28$ GeV in the $\sigma(pp\rightarrow bbA \rightarrow bb\mu\mu)$ events.

\end{abstract}
\end{titlepage}

\section{Introduction}
\label{ch:introduction}

The precise analyses \cite{Khachatryan:2015cwa,ATLAS:2014aga,Aad:2015ona,Chatrchyan:2013iaa,Sirunyan:2017dgc,ATLAS:2017ovn,ATLAS:2018gcr} have reported deviations from the Standard Model (SM) predictions which \cite{Sirunyan:2018kst} leave a window for new physics and point out the probability of the existence of extra Higgs bosons \cite{Aad:2019zwb,ATLAS:2019jzx,CMS:2019kjn,CMS:1900sig,Sirunyan:2019xls,Sirunyan:2019zdq,CMS:2019cid}. For instance, an excess at about 28 GeV in the invariant mass of the muon pairs is revealed by the CMS data obtained at both 8 TeV and 13 TeV \cite{Sirunyan:2018wim} COM energies which can arise from decays of some states which are not included in the SM. One of the most economical way is to extend the SM particle content by one Higgs doublet, which forms a class of THDMs \cite{Bhattacharyya:2015nca}. Since this Higgs doublet is allowed to couple to the SM particles it leads to some possible signal processes that can be tested at the collider experiments, and one of the salient features of THDMs is that one can realize light Higgs bosons consistent with the current experimental results \cite{Aggleton:2016tdd,Chun:2018vsn}. If a scalar state is realized as light as about 28 GeV, it can also be a viable candidate for the state which leads to the excess in the invariant mass of the muon pair.

In our study, we assume the dimuon excess arises from the presence of a light scalar state. Among a variety of different $Z_{2}$ symmetries, one of them can yield a class of THDMs in which the Higgs doublets distinguish the leptons so that one of the Higgs doublet can couple to the leptons at tree-level, while the other is not allowed \cite{Han:2021gfu,Nomura:2019wlo,Han:2018znu,Wang:2018hnw,Hashemi:2018kct}. We will consider this class of THDMs which is called Lepton Specific THDM (LS-THDM), and discuss its mass spectrum for viability of light Higgs bosons in light of several experimental constraints. 

\section{Lepton Specific THDM}
\label{sec:model}

The rich vacuum structure of THDMs can also yield some implications on charged lepton flavor violation (cLFV) or flavor changing neutral currents (FCNC). The severe constraints on the tree-level (cLFV) \cite{Adam:2013mnn,Aubert:2009ag} strongly restrict the Yukawa couplings between the charged leptons and the Higgs bosons. These constraints and the absence of FCNC processes can be satisfied by imposing a $\mathcal{Z}_{2}$ symmetry \cite{Ko:2012hd,Nomura:2019wlo}. We choose to impose a $\mathcal{Z}_{2}$ symmetry such that the scalar potential and the Yukawa Lagrangian remain invariant when the fields transform as $\phi_{1} \rightarrow -\phi_{1}$, $\phi_{2}\rightarrow \phi_{2}$ and $e_{R}\rightarrow -e_{R}$, where $e_{R}$ denotes the right-handed charged leptons. Assuming all other fields are even under this $\mathcal{Z}_{2}$ transformation leads to the following Yukawa Lagrangian:
\begin{equation}
\mathcal{L}_{Y} = -Y^{u}\overline{Q}\tilde{\phi}_{2}u_{R}+Y^{d}\overline{Q}\phi_{2}d_{R}+Y^{e}\overline{L}\phi_{1}e_{R}~,
\label{eq:YukawaLag}
\end{equation}

In LS-THDM, the large top quark mass \cite{Group:2009ad}, requires the SM-like Higgs boson to be formed mostly by $\phi_{2}$, while the light mass eigenstates of scalars are expected to arise from $\phi_{1}$. In this case, the couplings between the light scalars and the quarks become inversely proportional to $\tan\beta$. As summarized in Table \ref{tab:yuk}, $A-$ boson also negligibly couples to the quarks for large $\tan\beta$, while its coupling to the leptons is enhanced. Note that Table 1 assumes $h_{1}$ to be the SM-like Higgs boson, where $\alpha_{H}$ defines the mixing angle between the CP-even Higgs bosons as in Refs.\cite{Carena:2014nza,Hou:2017hiw}, while $\sin\beta$ and $\cos\beta$ are obtained from $\tan\beta \equiv v_{2}/v_{1}$. The couplings should be switched between $h_{1}$ and $h_{2}$, when $h_{2}$ is the SM-like Higgs boson. 
\begin{table}[h!]
\centering
\setstretch{1.5}
\begin{tabular}{|l|l|l|l|l|}
\hline
$Y_{u}^{h_{1}}, Y_{d}^{h_{1}}$,$Y_{l}^{h_{1}}$ & $\cos\alpha_{H}/\sin\beta$ & $\cos\alpha_{H}/\sin\beta$ & $-\sin\alpha_{H}/\cos\beta$  \\ \hline
$Y_{u}^{h_{2}}, Y_{d}^{h_{2}}$,$Y_{l}^{h_{2}}$ & $\sin\alpha_{H}/\sin\beta$ & $\sin\alpha_{H}/\sin\beta$ & $\cos\alpha_{H}/\cos\beta$  \\ \hline
$Y_{u}^{A}, Y_{d}^{A}$,$Y_{l}^{A}$ & $\cot\beta$& $-\cot\beta$& $\tan\beta$ \\ \hline
\end{tabular}
\caption{Effective Yukawa couplings between the quarks, leptons and the Higgs bosons.}
\label{tab:yuk}
\end{table}

The CP conservation and $\mathcal{Z}_{2}$ invariance yield the following scalar potantial in the usual notation \cite{Branco:2011iw,Diaz:2002tp,Chakraborty:2015raa,Mahmoudi:2009zx,Ivanov:2015nea}:
\begin{equation}
\setstretch{2.2}
\begin{array}{ll}
V(\Phi_{1},\Phi_{2}) & = m_{1}^{2}\Phi_{1}^{\dagger}\Phi_{1}+m_{2}^{2}\Phi_{2}^{\dagger}\Phi_{2}+m_{3}^{2}(\Phi_{1}^{\dagger}\Phi_{2}+\Phi_{2}^{\dagger}\Phi_{1}) \\
&+\dfrac{1}{2}\lambda_{1}(\Phi_{1}^{\dagger}\Phi_{1})^{2} +\dfrac{1}{2}\lambda_{2}(\Phi_{2}^{\dagger}\Phi_{2})^{2}+\lambda_{3}(\Phi_{1}^{\dagger}\Phi_{1})(\Phi_{2}^{\dagger}\Phi_{2}) \\
& +\lambda_{4}(\Phi_{1}^{\dagger}\Phi_{2})(\Phi_{2}^{\dagger}\Phi_{1}) +\left[\dfrac{1}{2}\lambda_{5}(\Phi_{1}^{\dagger}\Phi_{2})^{2}+{\rm h.c.}\right]~,
\end{array}
\label{eq:scalarpot}
\end{equation}
where $m_{1,2,3}$ are the mass terms, and $\lambda_{i}$ stand for the couplings between the Higgs doublets. Note that the terms with $m_{3}$, $\lambda_{3}$ and $\lambda_{5}$ softly break the $\mathcal{Z}_{2}$ symmetry required by the perturbativity condition \cite{Gunion:2002zf}.

After the electroweak symmetry breaking, the existence of a massless CP-odd Goldstone boson requires either $\sin\beta_{H}=0$ or $\cos\beta_{H}=0$, where $\beta_{H}$ is the mixing angle in the CP-odd Higgs sector which rotates the mixing away. Setting $\cos\beta_{H}=0$ leads to a massive CP-odd Higgs boson in the spectrum which is formed by $\Phi_{1}$, and its mass is
\begin{equation}
m_{A}^{2} = \dfrac{-m_{3}^{2}}{\sin\beta\cos\beta}-\lambda_{5}v^{2}~.
\label{eq:mA}
\end{equation}
where $v\equiv \sqrt{v_{1}^{2}+v_{2}^{2}}\simeq 246$ GeV to be consistent with the fermion and gauge boson masses. The charged Higgs boson mass in terms of $m_{A}$ is
\begin{equation}
m^{2}_{H^{\pm}} =  m^{2}_{A}-\dfrac{1}{2}(\lambda_{4}-\lambda_{5})v^{2}~.
\end{equation}
Note that $\lambda_{4}$ leads, in general, to lighter charged Higgs bosons, and thus it is assumed to be zero in our scan to realize heavier charged Higgs bosons. The masses for the CP-even Higgs bosons can be obtained by rotating the mixing parameters away by $\alpha_{H}$ in the mass-squared matrix yields the following masses from lighter to heavier:
\begin{equation}
\setstretch{1.2}
\begin{array}{rl}
m_{h_{1}}^{2}= & m_{A}^{2}\cos^{2}(\beta-\alpha_{H}) \\ & + v^{2}\biggr[\lambda_{1}\cos^{2}\beta\sin^{2}\alpha_{H} + \lambda_{2}\sin^{2}\beta\cos^{2}\alpha_{H} \\ &
 -\dfrac{1}{2}(\lambda_{3}+\lambda_{5})(\sin^{2}(\beta +\alpha_{H})-\sin^{2}(\beta -\alpha_{H})) \\ & +\left. \lambda_{5}\cos^{2}(\beta -\alpha_{H})\right]~, \\ & \\
m^{2}_{h_{2}} = & m_{A}^{2}\sin^{2}(\beta -\alpha_{H}) \\ & +v^{2}\biggr[ \lambda_{1}\cos^{2}\beta \cos^{2}\alpha_{H}+\lambda_{2}\sin^{2}\beta \sin^{2}\alpha_{H} \\ &+\dfrac{1}{2}(\lambda_{3}+\lambda_{5})(\sin^{2}(\beta +\alpha_{H})-\sin^{2}(\beta -\alpha_{H})) \\ & +\left. \lambda_{5}\sin^{2}(\beta -\alpha_{H})\right]~,\end{array}
\end{equation}

Our discussion about the masses and mixing in the Higgs sector is based on the tree-level calculations, and it cannot be completed without considering the contributions at loop levels. These contributions can be obtained by adding the radiative corrections to the scalar potential through the following equation \cite{Ferreira:2015pfi}:
\begin{equation}
V_{loop} = \dfrac{1}{64\pi^{2}}\displaystyle \sum_{\alpha}n_{\alpha}m_{\alpha}^{4}\left[\log\left(\dfrac{m_{\alpha}^{2}}{\mu^{2}} \right) - \dfrac{3}{2}\right]
\label{eq:loopV}
\end{equation}
where $\mu$ is the renormalization scale, $m_{\alpha}$ is the masses of particles contributing at loop level and $\alpha$ runs over all the particles that couple to the scalars at tree-level. $s_{\alpha}$ denotes the spin of the particles. In addition \cite{Ferreira:2015pfi}, 
\begin{equation}
n_{\alpha} = (-1)^{2s_{\alpha}}Q_{\alpha}C_{\alpha}(2s_{\alpha}+1)
\end{equation}
with
\begin{equation}
Q_{\alpha}= \left\lbrace \begin{array}{ll}
1 & {\rm for ~ neutral ~ particles~,} \\
2 & {\rm for ~ charged ~ particles~.}
\end{array}    \right.
\end{equation}
and $C_{\alpha}=3 (1)$ for quarks (leptons).

\section{Scanning Procedure and Experimental Constraints}
\label{sec:scan}

We have employed SPheno 4.0.3 \cite{Porod:2003um,Porod:2011nf} generated by SARAH 4.13.0 \cite{Staub:2015kfa,Staub:2010jh}, in our scans for LS-THDM. We follow a coupling-based scan spanned by the following parameters:
\begin{equation}
\begin{array}{llll}
-1 \leq & \lambda_{i} & \leq 1 & (i=1,2,3,5)~, \\
-25 \leq & m_{3}^{2} & \leq 25~{\rm TeV}^{2}~, & \\
1.2 \leq & \tan\beta & \leq 60~.&
\end{array}
\label{eq:params}
\end{equation}
After generating data we employ SusHi 1.7.0 \cite{Harlander:2012pb,Harlander:2016hcx,Bahl:2018zmf,Bahl:2019ago} to calculate the Higgs boson productions. Note that SusHi employs the parton distribution functions (pdf) which are provided by LHAPDF6 \cite{Buckley:2014ana,Skipis:2003aaa,Bourilkov:2006cj}. We set the pdf to be CTEQ6L1 \cite{Pumplin:2002vw} to compare our results with those from the experimental analyses. Note that $m_{1}^{2}$ and $m_{2}^{2}$ are calculated by the tadpole equations. We allow only solutions in our scans which are consistent with the electroweak symmetry breaking ($v^{2}= v_{1}^{2}+v_{2}^{2}\simeq v_{{\rm SM}}^{2}$). The ranges for the couplings are determined so that the spectra involve light Higgs bosons. We also allow negative values for these couplings; however, we require the solutions to be consistent with the condition of stability of the scalar potential. As discussed before, the large top quark mass requires $v_{2}\lesssim v_{{\rm SM}}$. Thus the stability condition restricts rather  coupling of $\Phi_{2}$ as $\lambda_{2} > 0$. In this context, the CP-odd Higgs ($A$) and the lighter CP-even Higgs ($h_{1}$) bosons are allowed to be light when they are formed mostly by $\phi_{1}$.

A problem with light scalar states arise from the rare $B-$meson decays such as $B_{s}\rightarrow \mu^{+}\mu^{-}$ and $B_{s}\rightarrow X_{s}\gamma$ where $X_{s}$ denotes a meson with a strange quark. These decay modes receive contributions from the scalar states \cite{Mahmoudi:2012un,Arbey:2012ax,Nis:2017obl,Arnan:2017lxi}, and the agreement between the SM predictions \cite{Bobeth:2013uxa,Bobeth:2013tba,Hermann:2013kca} and the experimental measurements \cite{Aaij:2012nna,Amhis:2019ckw,Amhis:2012bh} yields a strong impact on the light scalar states. As summarized in Table \ref{tab:yuk}, $\tan\beta$ suppresses the coupling between the light scalars and quarks, which can yield consistent rare $B-$meson decays. We apply the following constraints to check if a solution is consistent with the Higgs boson mass constraints and rare $B-$meson decays:
\begin{equation}
\setstretch{1.1}
\begin{array}{lcl}
123 \leq & m_{h_{1}}~{\rm or}~m_{h_{2}} & \leq 127~{\rm GeV}~, \\
& m_{H^{\pm}} & \gtrsim 80~{\rm GeV}~, \\
0.8\times 10^{-9} \leq & {\rm BR}(B_s \rightarrow \mu^+ \mu^-) &
  \leq 6.2 \times10^{-9} ~ (2\sigma)~,\\
  2.99 \times 10^{-4} \leq &
  {\rm BR}(B_{s} \rightarrow X_{s} \gamma) &
  \leq 3.87 \times 10^{-4}~(2\sigma)~.
\end{array}
\label{eq:constraints}
\end{equation}
\section{Higgs Mass Spectrum}
\label{sec:masses}
\begin{figure}[ht!]
\centering
\subfigure{\includegraphics[scale=0.35]{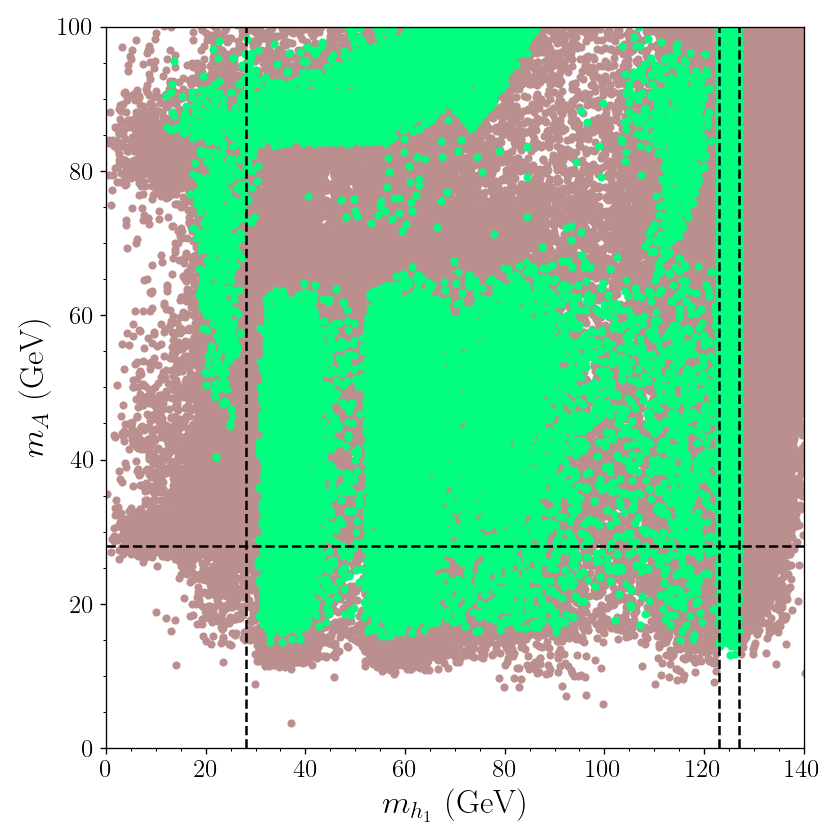}}%
\subfigure{\includegraphics[scale=0.35]{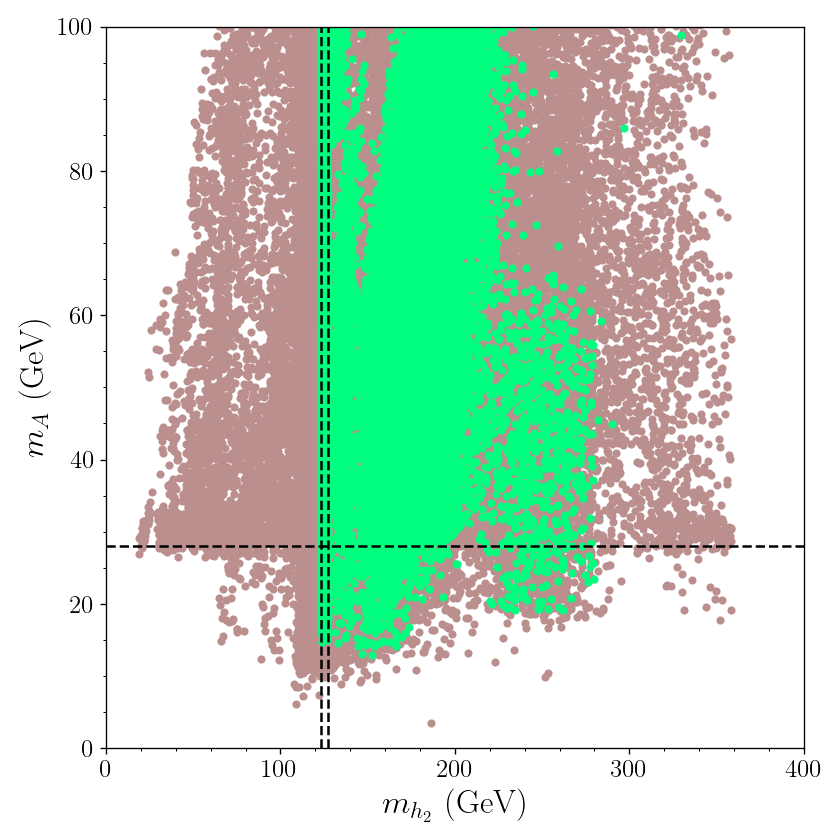}}
\subfigure{\includegraphics[scale=0.35]{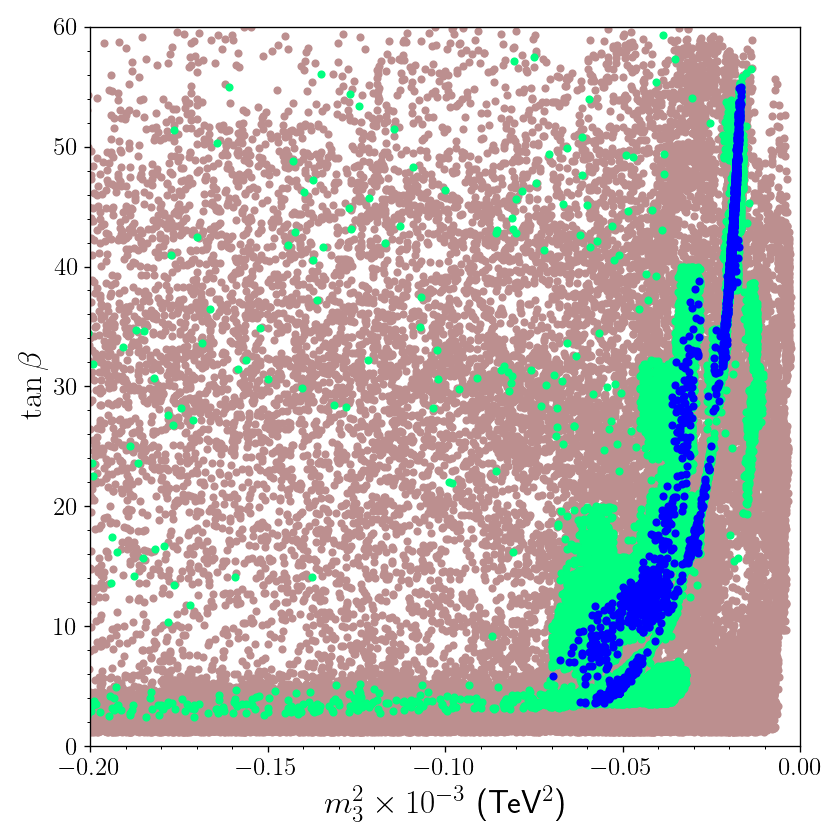}}%
\subfigure{\includegraphics[scale=0.35]{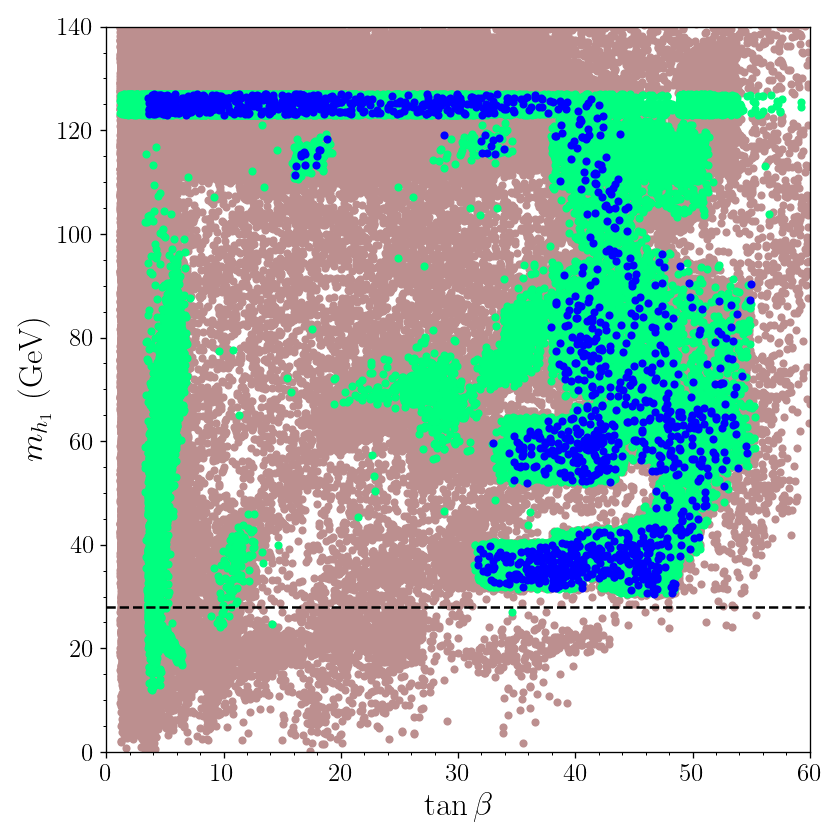}}
\caption{The Higgs boson masses and the fundamental parameters in the $m_{A}-m_{h_{1}}$, $m_{A}-m_{h_{2}}$, $\tan\beta - m_{3}^{2}$ and $m_{h_{1}}-\tan\beta$ planes. Green points satisfy the constraints from the Higgs boson masses and rare $B-$meson decays. The blue points in the bottom planes form a subset of green and represent the solutions in which $26 \leq m_{A} \leq 30$ GeV. The horizontal line indicate the region where the plotted Higgs boson mass is 28 GeV. In the $m_{A}-m_{h_{1}}$ plane the first vertical line shows the region of $m_{h_{1}}=28$ GeV, while the latter two vertical lines show the solutions with $m_{h_{1}}=123$ GeV and $m_{h_{1}}=127$ GeV respectively.}
\label{fig:Hmasses}
\end{figure}

In this section we discuss the Higgs boson mass spectrum and parameter space which yields light Higgs bosons in terms of the fundamental parameters as shown in Figure \ref{fig:Hmasses} with plots in the $m_{A}-m_{h_{1}}$, $m_{A}-m_{h_{2}}$, $\tan\beta - m_{3}$ and $m_{h_{1}}-\tan\beta$ planes. The $m_{A}-m_{h_{1}}$ plane shows that a light CP-odd Higgs boson ($m_{A} \lesssim 100$ GeV) can be widely realized, and such solutions can predict also a light CP-even Higgs boson ($m_{h_{1}} \gtrsim 10$ GeV) without violating any constraint listed in Eq.(\ref{eq:constraints}). If the spectrum involves a light CP-even Higgs boson, the $h_{2}$ is required to be the SM-like Higgs boson, and these light CP-even Higgs boson solutions correspond to the green points between the two vertical dashed lines in the $m_{A}-m_{h_{2}}$ plane. However, realizing light CP-odd Higgs boson is also possible for $m_{h_{1}} \simeq 125$ GeV, for which $m_{h_{2}}$ can be as heavy as about 300 GeV. The consistent electroweak symmetry breaking restricts $m_{3}^{2}$ to be negative in all physical solutions as shown in the $\tan\beta-m_{3}^{2}$ plane. The light CP-odd higgs boson condition (blue points) restricts $m_{3}^{2}$ 
as $m_{3}^{2} \gtrsim -0.07$ TeV$^{2}$. In addition, $\tan\beta$ can be as low as about 2 for the largest negative $m_{3}$ value, and it can increase to about 55 when the negative $m_{3}^{2}$ is low. The correlation between $\tan\beta$ and $m_{3}^{2}$, which yields $m_{A} \sim 28$ GeV is a direct consequence of Eq.(\ref{eq:mA}). It is also possible to have light CP-odd and CP-even Higgs bosons simultaneously in the spectra as shown with blue points in the $m_{h_{1}}-\tan\beta$ plane. These solutions can be realized when $\tan\beta \gtrsim 30$.  
\begin{figure}[ht!]
\centering
\subfigure{\includegraphics[scale=0.35]{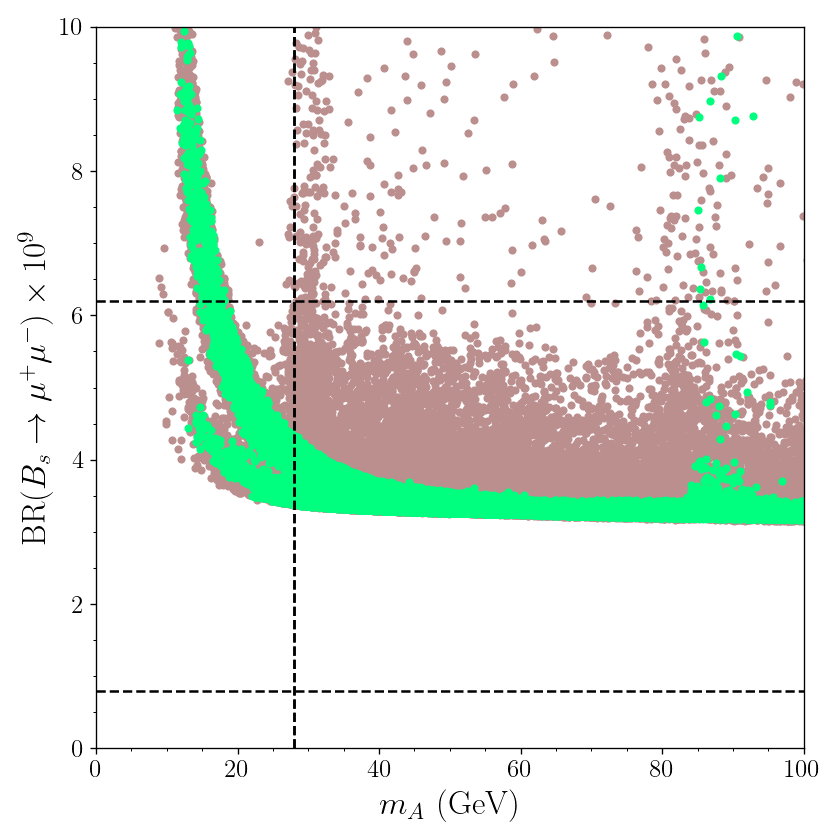}}%
\subfigure{\includegraphics[scale=0.35]{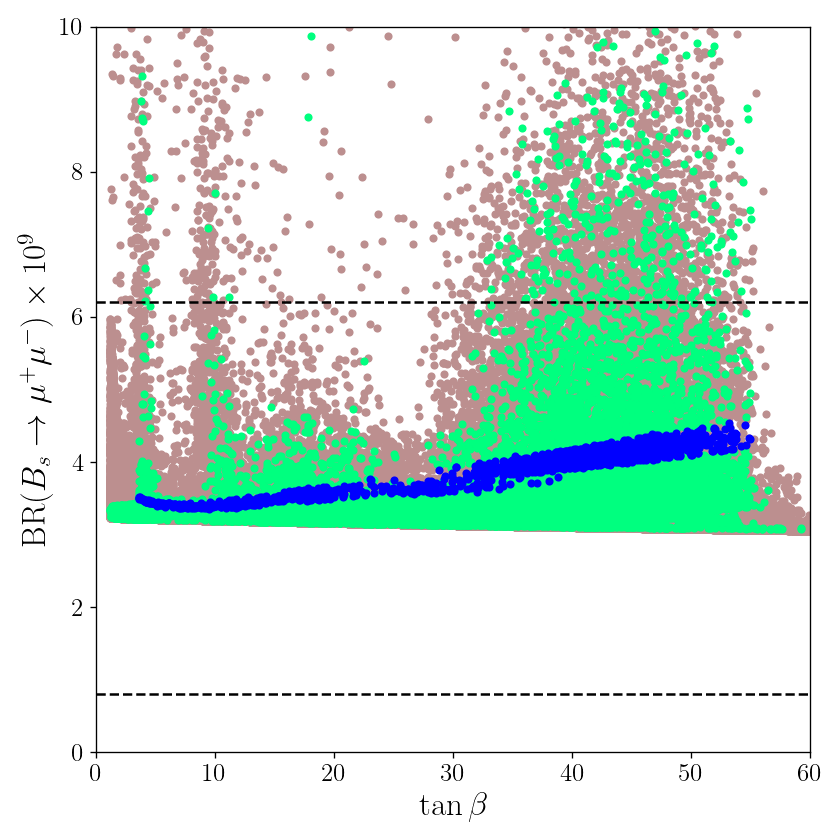}}
\subfigure{\includegraphics[scale=0.35]{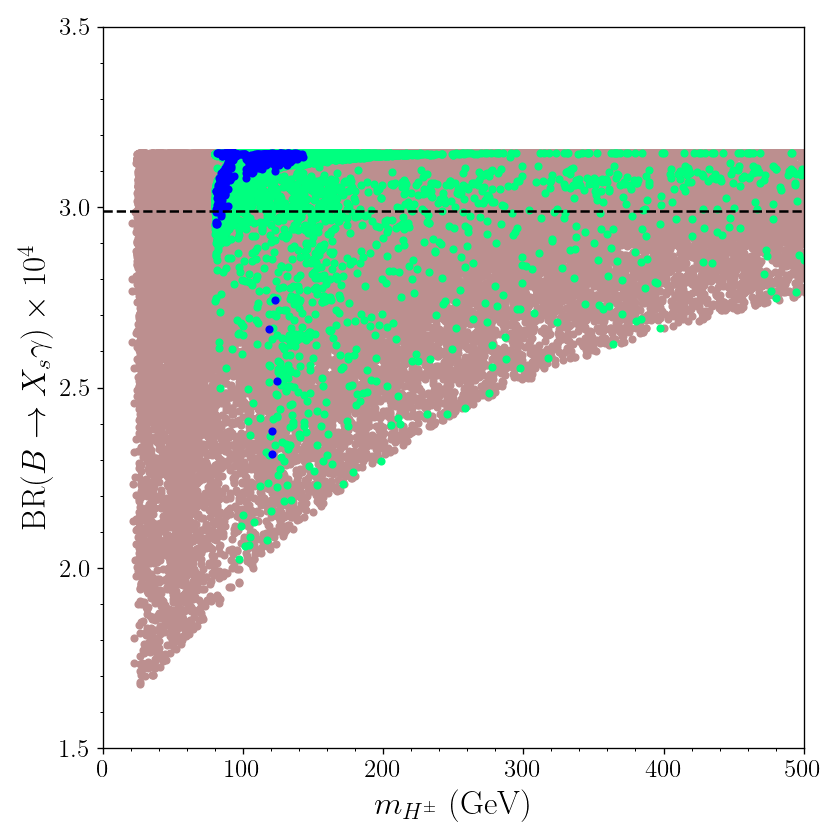}}%
\subfigure{\includegraphics[scale=0.35]{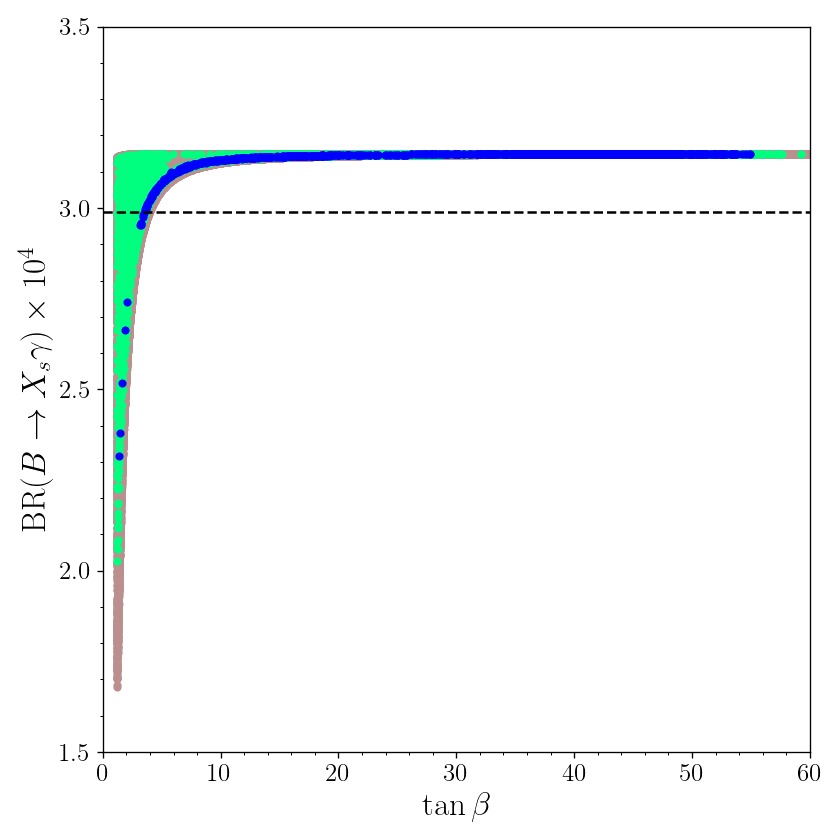}}
\caption{Impact from the rare $B-$meson decays in the ${\rm BR}(B_{s}\rightarrow \mu^{+}\mu^{-})-m_{A}$, ${\rm BR}(B_{s}\rightarrow \mu^{+}\mu^{-})-\tan\beta$, ${\rm BR}(B_{s}\rightarrow X_{s}\gamma)-m_{H^{\pm}}$ and ${\rm BR}(B_{s}\rightarrow X_{s}\gamma)-\tan\beta$ planes. The color coding is the same as Figure \ref{fig:Hmasses}, except the constraints from the plotted $B-$meson decays are not applied to the green points. The allowed ranges for these decays are shown with horizontal dashed lines.}
\label{fig:bdecays}
\end{figure}

We plot ${\rm BR}(B_{s}\rightarrow \mu^{+}\mu^{-})$ (top) and ${\rm BR}(B_{s}\rightarrow X_{s}\gamma)$ (bottom) decay modes in Figure \ref{fig:bdecays} with plots in the ${\rm BR}(B_{s}\rightarrow \mu^{+}\mu^{-})-m_{A}$, ${\rm BR}(B_{s}\rightarrow \mu^{+}\mu^{-})-\tan\beta$, ${\rm BR}(B_{s}\rightarrow X_{s}\gamma)-m_{H^{\pm}}$ and ${\rm BR}(B_{s}\rightarrow X_{s}\gamma)-\tan\beta$ planes. As seen from the top planes, ${\rm BR}(B_{s}\rightarrow \mu^{+}\mu^{-})$ strongly depends on the mass of the CP-odd Higgs boson mass, while it is possible to realize consistent solutions with $m_{A}\simeq 28$ GeV. In these light CP-odd Higgs boson solutions, one can observe a slight enhancement with $\tan\beta$ as shown with the blue points in the ${\rm BR}(B_{s}\rightarrow \mu^{+}\mu^{-})-\tan\beta$. However, this enhancement does not lead to inconsistently large values for ${\rm BR}(B_{s}\rightarrow \mu^{+}\mu^{-})$. In contrast to the $B_{s}\rightarrow \mu^{+}\mu^{-}$ decay mode, $B_{s}\rightarrow X_{s}\gamma$ seems to yield a stronger impact on the LS-THDM implications. As shown in the bottom planes of Figure \ref{fig:bdecays}. Extra contributions to this decay mode arise from the diagrams involving the charged Higgs boson, but the ${\rm BR}(B_{s}\rightarrow X_{s}\gamma)-m_{H^{\pm}}$ plane shows that it is possible to realize consistent solutions even when $m_{H^{\pm}}\gtrsim 80$ GeV. Indeed, if one requires $m_{A}\sim 28$ GeV (blue points), then $m_{H^{\pm}}$ is bounded at about 150 GeV from above consistent with the experimental measurements on ${\rm BR}(B_{s}\rightarrow X_{s}\gamma)$. However, the results reveal that most of the parameter space is excluded by the ${\rm BR}(B_{s}\rightarrow X_{s}\gamma)$ constraint, and the $\tan\beta$ parameter is strongly effective in determining the consistent ${\rm BR}(B_{s}\rightarrow X_{s}\gamma)$ values. Indeed, the ${\rm BR}(B_{s}\rightarrow X_{s}\gamma)-\tan\beta$ plane reveals almost a unique correlation between ${\rm BR}(B_{s}\rightarrow X_{s}\gamma)$ and $\tan\beta$.

\begin{figure}[ht!]
\centering
\subfigure{\includegraphics[scale=0.45]{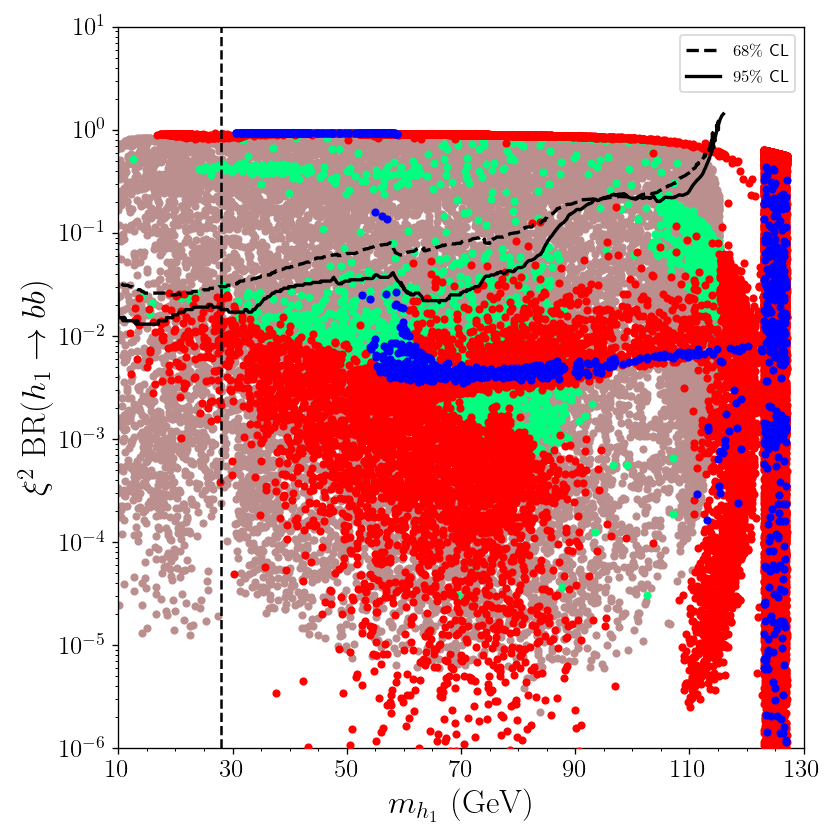}}%
\subfigure{\includegraphics[scale=0.45]{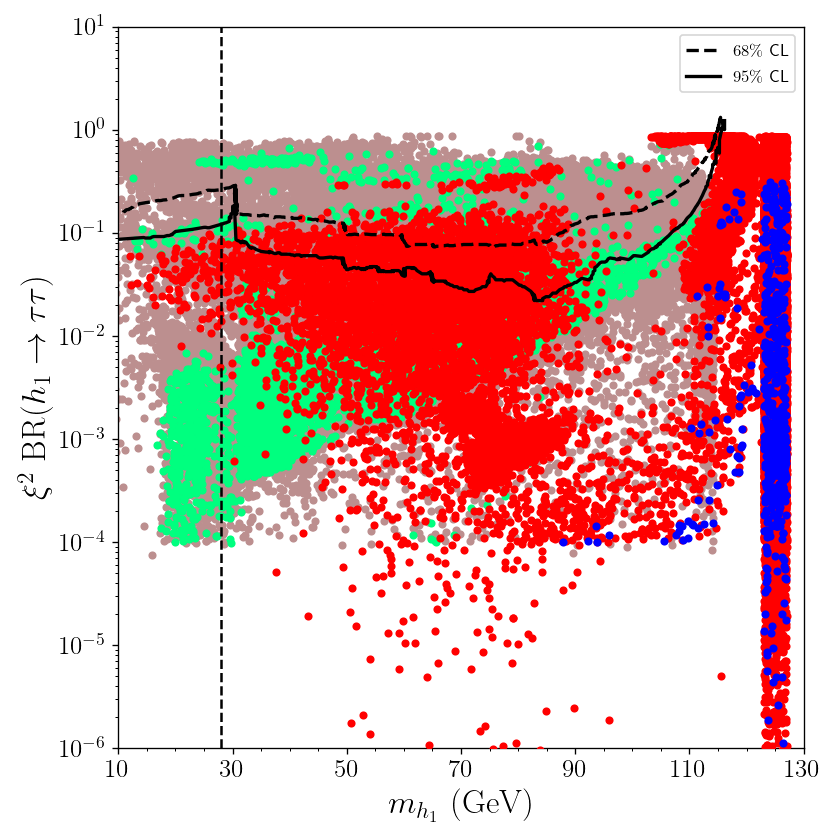}}
\caption{LEP bounds on the CP-even Higgs boson through $h_{1}\rightarrow bb$ (left) and $h_{1}\rightarrow \tau\tau$ (right) decay modes. The solid and dashed curves represent the exclusion curves from the LEP data \cite{Barate:2003sz} at the $95\%$ and $68\%$ confidence levels (CLs) respectively. All points are consistent with the electroweak symmetry breaking and fermion masses. Green points satisfy the constraints from the Higgs boson masses and rare $B-$meson decays. Red points form a subset of green and they are consistent with the LEP bounds. Note that the bound on $h_{1}\rightarrow bb$ is not applied in the left panel, while that on $h_{1}\rightarrow \tau\tau$ is not included in the right panel. Furthermore, blue points are a subset of red and they represent the solutions with $26 \leq m_{A} \leq 30$ GeV. The vertical dashed lines show the solutions with $m_{h_{1}} = 28$ GeV, and $\xi =g_{hZZ}/g_{hZZ}^{{\rm SM}}$ denotes the coupling between the Higgs and $Z$ bosons normalized to its SM value.}
\label{fig:LEP}
\end{figure} 

In addition to rare $B-$meson decays, the light Higgs bosons can be constrained by the results from thorough analyses over the LEP data \cite{Barate:2003sz}. We display the impact from the LEP data on the light $h_{1}$ solutions in Figure \ref{fig:LEP} through $h_{1}\rightarrow bb$ (left) and $h_{1}\rightarrow \tau\tau$ (right) decay modes. The red points in both planes of Figure \ref{fig:LEP} show that the LEP data can allow the CP-even Higgs boson as light as about 10 GeV. However, if one searches for light CP-odd Higgs boson solutions (blue), such solutions bound $m_{h_{1}}$ at about 55 GeV. The solutions for the spectra involving $m_{A}\sim m_{h_{1}} \simeq 28 $ GeV are excluded by the LEP data (blue points above the curves in the left panel). We also present the LEP bound for the ${\rm BR}(h_{1}\rightarrow \tau\tau) $ in the $\xi^{2}{\rm BR}(h_{1}\rightarrow \tau\tau)-m_{h_{1}}$ plane. As seen from the red and blue points, the impact from the ${\rm BR}(h_{1}\rightarrow \tau\tau)$ measurements is not as strong as that from the ${\rm BR}(h_{1}\rightarrow bb)$ decay mode, and many solutions can easily satisfy the constraint on the ${\rm BR}(h_{1}\rightarrow \tau\tau)$.

\begin{figure}[h!]
\centering
\subfigure{\includegraphics[scale=0.35]{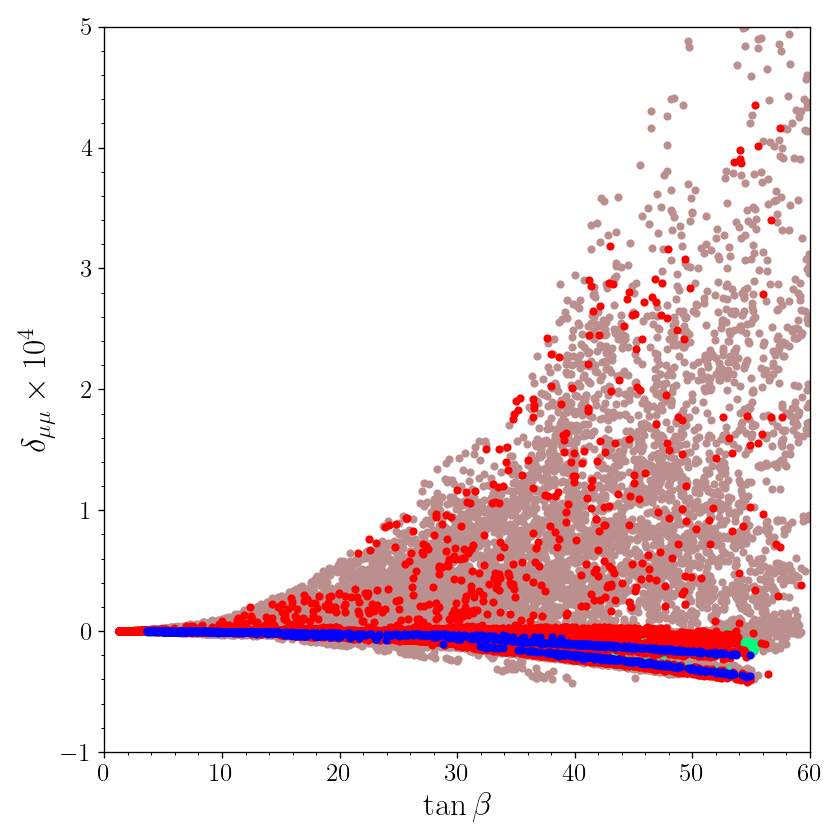}}%
\subfigure{\includegraphics[scale=0.35]{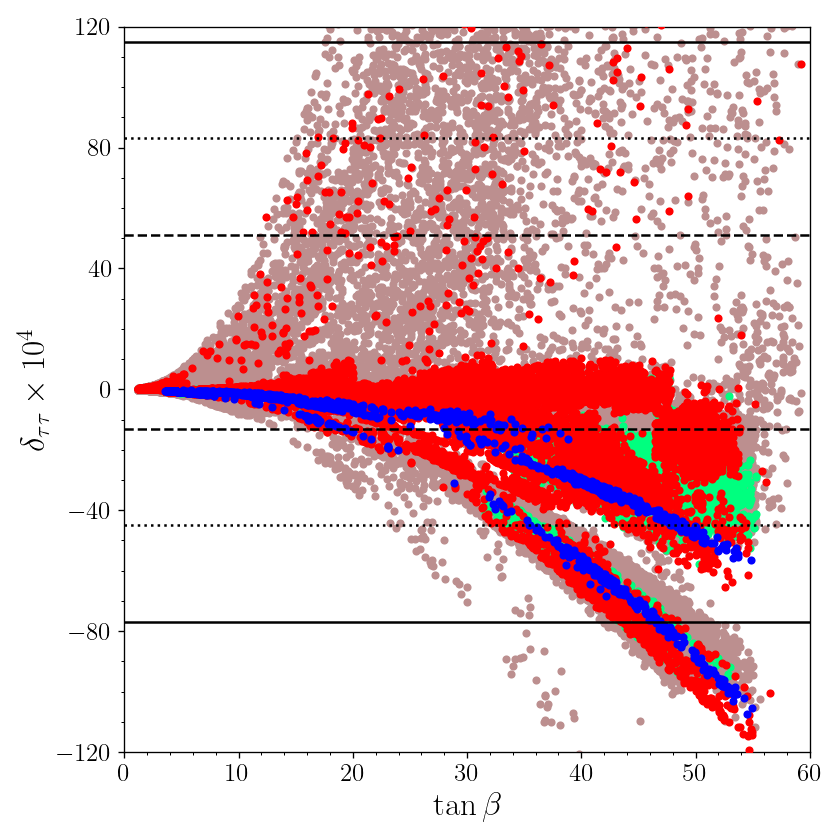}}
\subfigure{\includegraphics[scale=0.35]{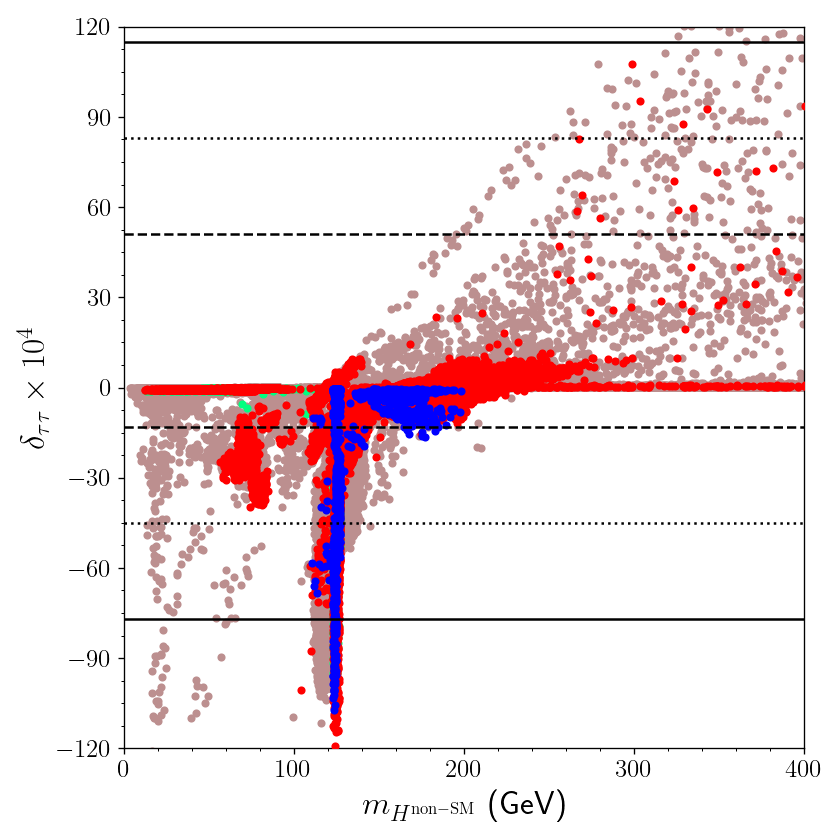}}%
\subfigure{\includegraphics[scale=0.35]{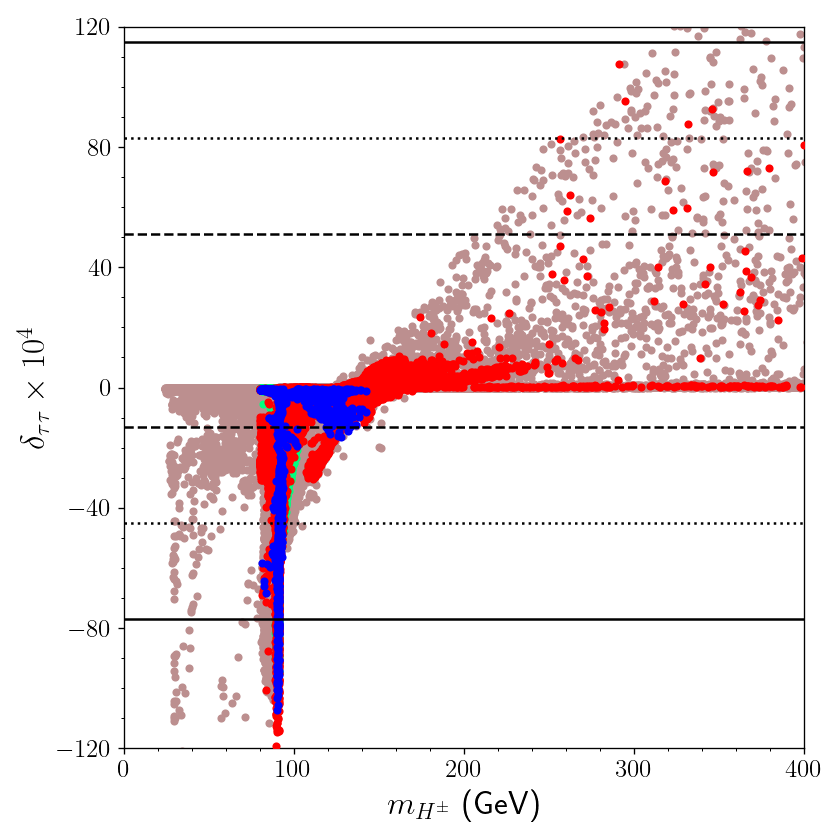}}
\caption{Plots for the deviation from the LFU in $\Gamma_{Z\rightarrow ll}$ decays, where $l=\mu,\tau$. All points are consistent with the electroweak symmetry breaking and fermion masses. Green points satisfy the constraints from Higgs boson mass and rare $B-$meson decays. The red points are also compatible with the LEP data. Blue points form a subset of red, and they represent the solutions with $26 \leq m_{A}\leq 30$ GeV. The horizontal lines represent the experimental bounds on LFU for $\delta_{\tau\tau}$  in top-right and bottom panels.}
\label{fig:LFUZ}
\end{figure}

Even though the LEP bound on $h\rightarrow \tau\tau$ is relatively weaker, the leptonic decays of the Higgs bosons can also be constrained by considering the Lepton Flavor Universality (LFU). Even though LS-THDM leaves the symmetrical structure of the SM intact and the solutions are constrained by the effective values of the Yukawa couplings, the extra Higgs bosons can still yield a deviation from LFU at the loop level. Such a deviation is enhanced by $\tan\beta$ \cite{Abe:2015oca,Chun:2016hzs,Wang:2018hnw}, and the solutions with large $\tan\beta$ can receive  strong impact from the experimental constraints on LFU \cite{ALEPH:2005ab}. We display our results in Figure \ref{fig:LFUZ} for the deviation in LFU within LS-THDM framework in correlation with $\tan\beta$ (top panels) and the masses of the Higgs bosons (bottom panels) which are not included in the SM. The color coding is the same as in Figure 3. The horizontal dashed, dotted, solid lines in $\delta_{\tau\tau}$ plots represent the bounds within $1\sigma$, $2\sigma$ and $3\sigma$ respectively. The definitions of $\delta_{\mu\mu}$ and $\delta_{\tau\tau}$ and the experimental constraints within $1\sigma$ can be expressed as follows:

\begin{equation}
\dfrac{\Gamma(Z\rightarrow \mu\mu)}{\Gamma(Z\rightarrow ee)} = 1 + \delta_{\mu\mu} \simeq 1.0009 \pm 0.0028~,\hspace{0.3cm}
\dfrac{\Gamma(Z\rightarrow \tau\tau)}{\Gamma(Z\rightarrow ee)} = 1 + \delta_{\tau\tau}\simeq 1.0019\pm 0.0032~.
\label{eq:deviation}
\end{equation}

The top-left panel of Figure \ref{fig:LFUZ} shows that $\delta_{\mu\mu} \sim 0$, while the deviation of $\delta_{\tau\tau}$ can exceed the $1\sigma$ range if $\tan\beta \gtrsim 20$ as shown in the top-right panel of Figure \ref{fig:LFUZ}. The bottom panels represent the correlation with the Higgs boson masses, where $H^{{\rm non-SM}}$ stands for the second CP-even Higgs boson, which is not assigned to be SM-like Higgs boson in LS-THDM. Even though the Higgs boson masses lead to a weaker suppression in the LFU deviation, comparing with the results in the $\delta_{\tau\tau}-\tan\beta$ plane, at least one of the Higgs bosons should have the mass greater than about  120 GeV to be compatible with the LFU constraints within $1\sigma$ when $\tan\beta \gtrsim 20$. As given in Table 1, the fermionic couplings of the extra Higgs bosons are weakened if $\tan\beta$ is large. Therefore a possible dimuon excess arising from one of these Higgs bosons can more likely be realized  for $\tan\beta \lesssim 20$, for which the deviation in $Z\rightarrow \tau\tau$ ($\delta_{\tau\tau}$) does not exceed the $1\sigma$ range set by the analyses. In this context, we will not consider LFU in the next section, unless we discuss the solutions with large $\tan\beta$. Note that we also checked the oblique parameters \cite{Pomarol:1993mu,Peskin:2001rw,Gerard:2007kn}, and  we observed a weak impact that a few solutions of light CP-odd Higgs boson can yield $\Delta T$ slightly lower than its $1\sigma$ bound set by the experimental fits \cite{Baak:2014ora} for $\tan\beta \gtrsim 20$.

Before concluding the experimental constraints employed in our analyses, one should consider a long term discrepancy between the SM predictions and the experimental measurements of the muon anomalous magnetic moment (muon $g-2$, for short). A possible solution for this discrepancy rather requires large $\tan\beta$ ($\gtrsim 30$, see for instance \cite{Wang:2018hnw}), and it leads to a fact that a dimuon excess cannot be realized simultaneously with the muon $g-2$ solution within the minimal LS-THDM framework. In this context, we do not require solutions to solve or improve the SM predictions on muon $g-2$, but we accept only solutions which do not worsen the SM predictions.

\section{Associated Production of CP-odd Higgs Boson and b-jets}
\label{sec:assocProd}
\begin{figure}[ht!]
\centering
\subfigure{\includegraphics[scale=0.35]{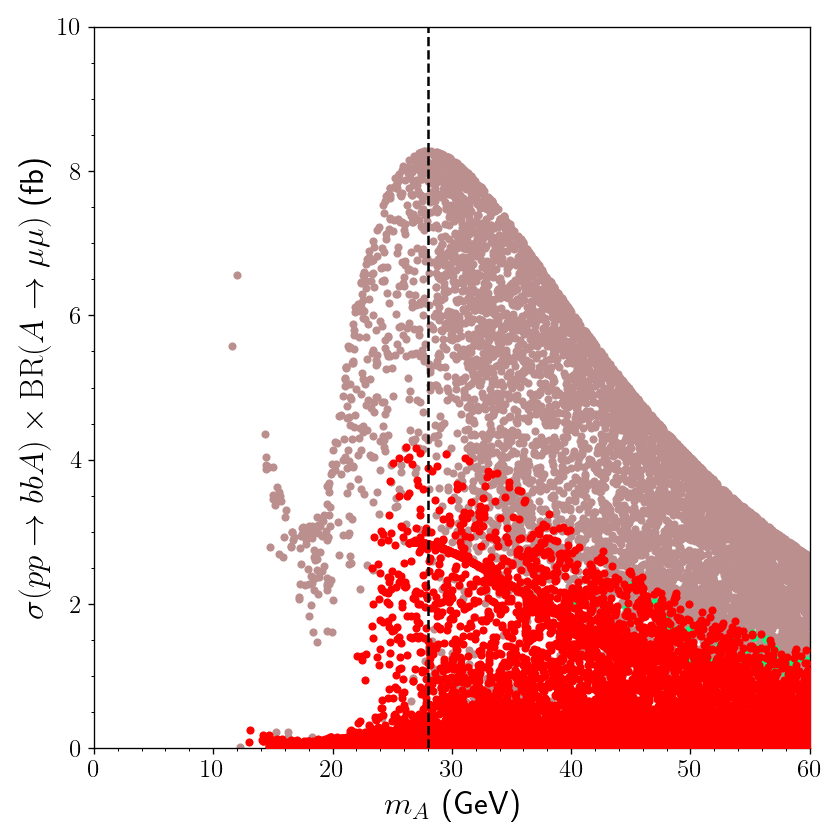}}%
\subfigure{\includegraphics[scale=0.35]{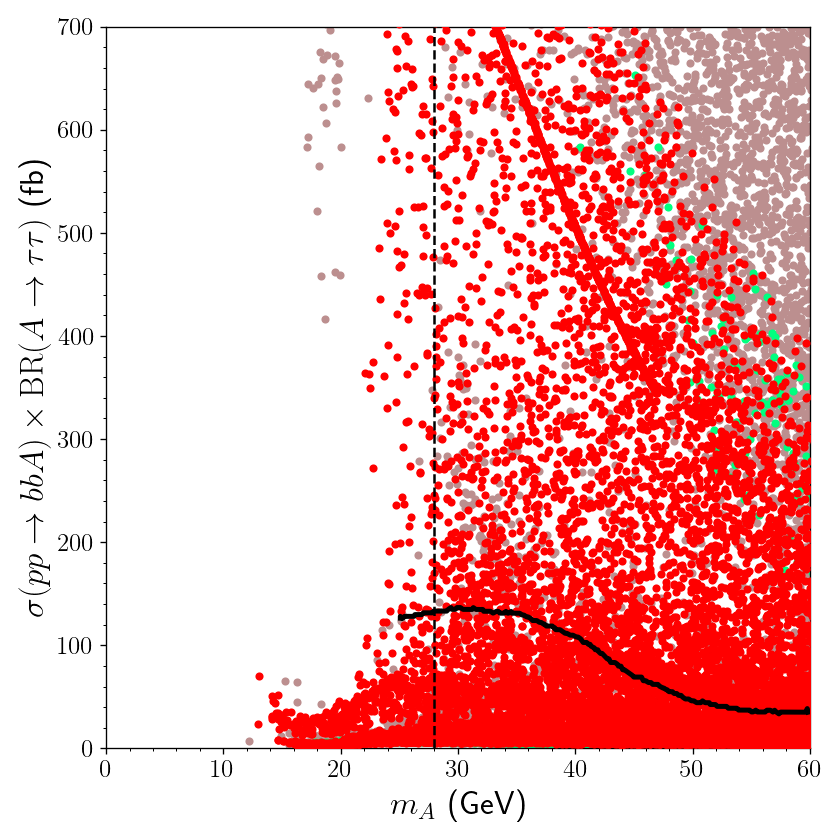}}
\subfigure{\includegraphics[scale=0.35]{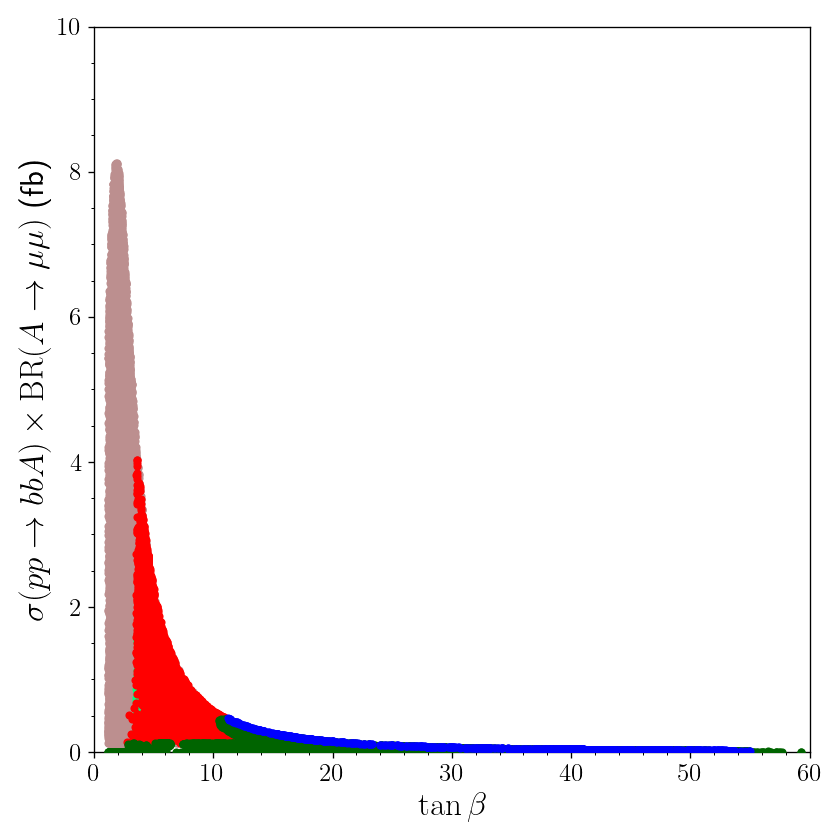}}%
\subfigure{\includegraphics[scale=0.35]{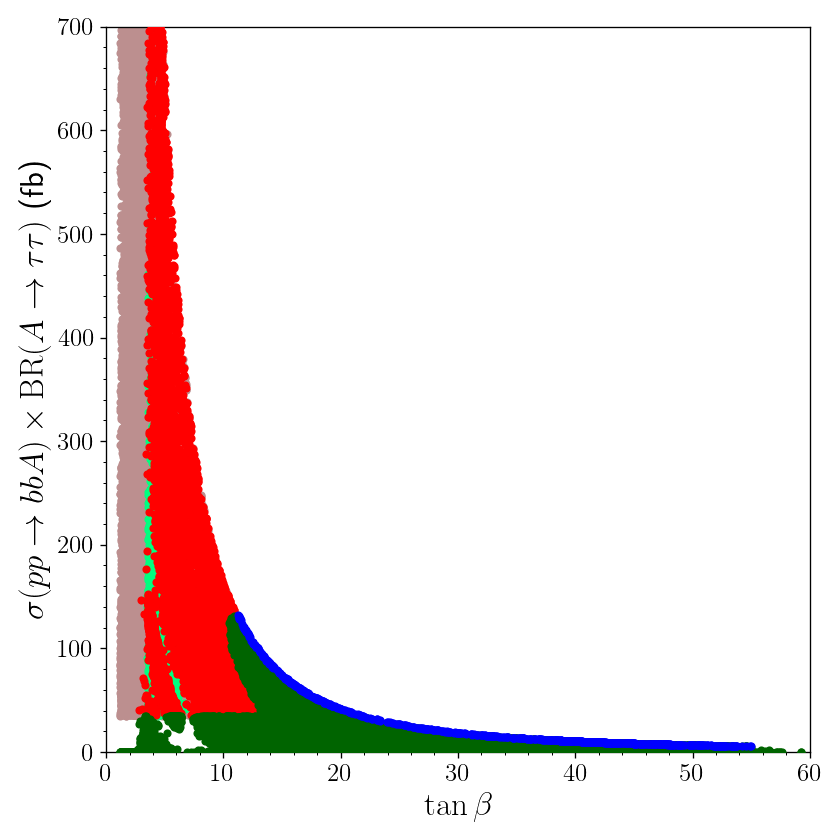}}
\caption{Cross-section values of the events $pp\rightarrow bbA \rightarrow bb\mu\mu$ (left) and $pp\rightarrow bbA \rightarrow bb\tau\tau$ (right) at 8 TeV COM in correlation with $m_{A}$ (top) and $\tan\beta$ (bottom). The top planes use the same color coding as Figure \ref{fig:LEP}. The vertical dashed lines in these planes show the solutions with $m_{A}=28$ GeV. The black curve in the top right plane represents the experimental bound from the analyses \cite{Sirunyan:2017uvf}. This bound allow the solutions shown in green in the bottom planes. In addition, the blue points form a subset of green and they indicate solutions for $26 \leq m_{A} \leq 30$ GeV.}
\label{fig:bbACMS}
\end{figure}

In previous section, we discussed the possible spectra involving light Higgs bosons and impact from several experimental results such as those from LEP and LHCb. However, the mass scales realized in our scans for the light Higgs bosons receive strong impacts also from several recent analyses. Such an impact can be stronger when the events involve the $A\rightarrow \tau\tau$ decay channel especially when the CP-odd Higgs boson is produced at LHC in association with a pair of bottom quarks \cite{Sirunyan:2017uvf}. We display results for the cross-section values obtained in our scans in Figure \ref{fig:bbACMS} for the events $pp\rightarrow bbA \rightarrow bb\mu\mu$ (left) and $pp\rightarrow bbA \rightarrow bb\tau\tau$ (right) at 8 TeV COM in correlation with $m_{A}$ (top) and $\tan\beta$ (bottom). The top planes use the same color coding as Figure \ref{fig:LEP}. The total cross-sections for the events are calculated by the following equations approximately:
\begin{equation}
\setstretch{1.5}
\begin{array}{ll}
\sigma(pp\rightarrow bbA \rightarrow bb\mu\mu) &\approx \sigma(pp\rightarrow bbA)\times {\rm BR}(A\rightarrow \mu\mu) \\
\sigma(pp\rightarrow bbA \rightarrow bb\tau\tau) &\approx \sigma(pp\rightarrow bbA)\times {\rm BR}(A\rightarrow \tau\tau)
\end{array}
\end{equation}

The bound on the cross-sections of the events which involved $A\rightarrow \mu\mu$ is observed to be about 200 fb for $m_{A}\simeq 28$ GeV, and it decreases to about 100 fb for relatively heavier CP-odd Higgs bosons \cite{Sirunyan:2017uvf}. However, as we show in the $\sigma(pp\rightarrow bbA)\times {\rm BR}(A\rightarrow \mu\mu)-m_{A}$ plane, LS-THDM can predict the cross-section only as large as about 4 fb. Indeed, this bound can yield a strong impact for the other classes of THDM in which the couplings between the CP-odd Higgs boson and the quarks are not suppressed by $\tan\beta$ \cite{Bernon:2014nxa}. On the other hand, the results from the analyses can be relevant to the implications of LS-THDM when the $A\rightarrow \tau\tau$ decay channel is involved. The top right panel of Figure \ref{fig:bbACMS} shows that most of the solutions can be excluded (point above the solid black curve) by the results reported in Ref.\cite{Sirunyan:2017uvf}. As shown in the bottom planes, the largest cross-section consistent with results on $A\rightarrow \tau\tau$ is realized when $\tan\beta \sim 10$. The cross-section rapidly decreases near to zero when $\tan\beta \gtrsim 20$.
\begin{figure}[ht!]
\centering
\subfigure{\includegraphics[scale=0.42]{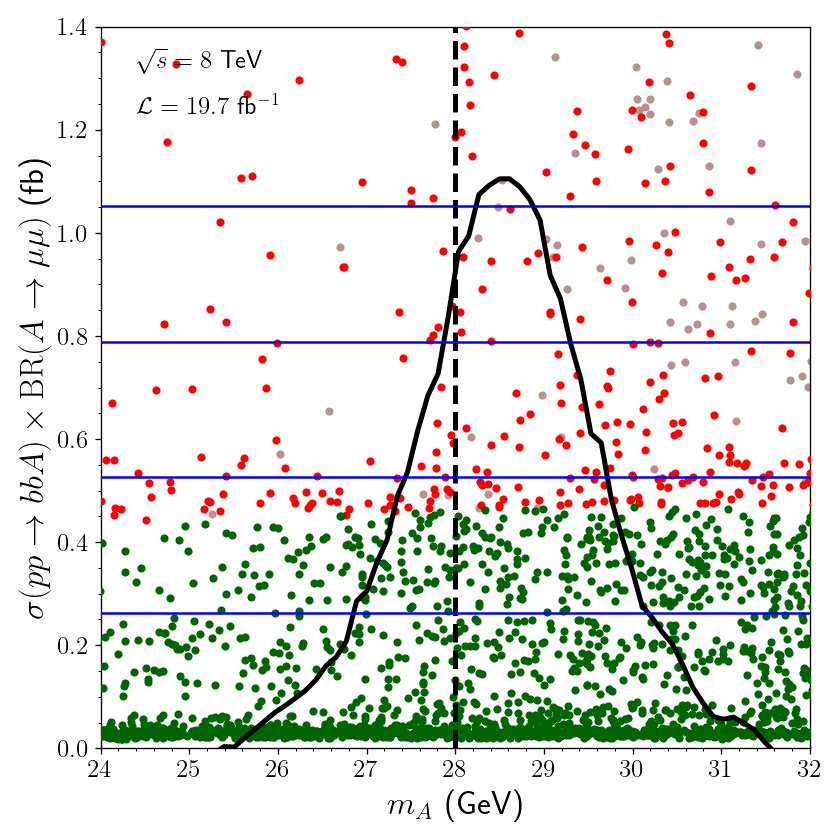}}%
\subfigure{\includegraphics[scale=0.42]{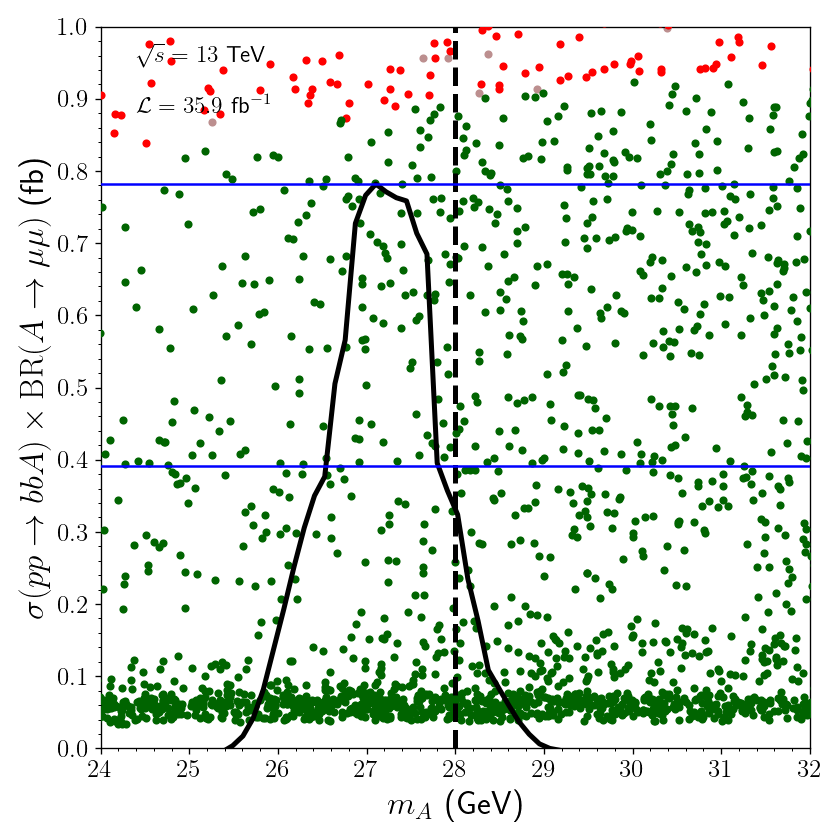}}
\caption{Plots in the $\sigma(pp\rightarrow bbA)\times {\rm BR}(A\rightarrow \mu\mu)-m_{A}$ planes for 8 TeV COM with 19.7 fb$^{-1}$ Luminosity (left) and 13 TeV with 35.9 fb$^{-1}$ Luminosity. The color coding is the same as the bottom panles of Figure \ref{fig:bbACMS}. The black curves show the observed cross-sections for the signal events \cite{Sirunyan:2018wim}. The blue horizontal lines show the standard deviations from $1\sigma$ to $4\sigma$ in the left panel and $1\sigma$ to $2\sigma$ in the right panel from bottom to top, respectively.}
\label{fig:sigmaexcess}
\end{figure}

The production of CP-odd Higgs boson in association with a pair of $b-$quarks takes special place, since recent analyses performed over the invariant mass of the muon pair in the final state have reported a significant excess at about $m_{\mu\mu}\simeq 28$ GeV. The analyses over the collisions at 8 TeV COM energy result in about $4.2\sigma$ deviation, while the analyses over 13 TeV COM-collisions have updated the excess to $2\sigma$ deviation from the background \cite{Sirunyan:2018wim}. Indeed, such an excess has been observed over a strong background which is mainly formed by the $t\bar{t}-$pair production, and Drell-Yan processes. Even though the vector boson fusion processes can take part, they yield an impact when $m_{\mu\mu}\sim 90$ GeV at which all the observed events seem to arise from the background processes only. We show the results that can accommodate such an excess for the mass scales $26 \lesssim m_{A} \lesssim 30$ GeV in the LS-THDM framework in Figure \ref{fig:sigmaexcess} for 8 TeV COM with 19.7 fb$^{-1}$ luminosity (left) and 13 TeV with 35.9 fb$^{-1}$ luminosity (right). Note that the curves are obtained only for the signal by subtracting the cross-section of the background processes from the total cross-section represented in Ref. \cite{Sirunyan:2018wim}. In our results (green and red points), we did not apply the cuts described in Ref. \cite{Sirunyan:2018wim}, thus our results include an uncertainty due to the missing factor of the kinetic efficiency. However, since more than $95\%$ of the signal events are expected to pass the selection rules in the considered kinematic region \cite{Sirunyan:2017uvf}, this uncertainty does not exceed $5\%$. As seen from the left panel of Figure \ref{fig:sigmaexcess}, LS-THDM can provide some solutions which can yield an excess phenomenologically at about $4\sigma$ (red points); however, these solutions are excluded by the bound from the $A\rightarrow \tau\tau$ bound discussed in Figure \ref{fig:bbACMS}. The allowed solutions (green points) can accommodate this excess up to about $2\sigma$ over the background at 8 TeV COM, which yield a cross-section of about 0.44 fb at most. On the other hand, the right panel shows that these solutions can lead to the observed excess resulted from the sensitive analyses for the 13 TeV-collisions. The peak shown with the black curve indicates a $2\sigma$ deviation, and we realize a number of green points around this peak.  Similarly, the solutions yielding a larger deviation (red points) are excluded by the bound on $A\rightarrow \tau\tau$. Comparing the results of Figure \ref{fig:sigmaexcess} with the bottom panels of Figure \ref{fig:bbACMS} one can read the $\tan\beta$ values for the excess. The solutions yielding about a $2\sigma$ excess at 8 TeV COM-collisions are realized when $\tan\beta \sim 11$ and the cross-section values decrease such that about a $1\sigma$ excess can be observed when $\tan\beta \sim 15$. Similarly, the solutions with $\tan\beta \sim 12$ predict about a $2\sigma$ excess, while those with $\tan\beta \sim 17$ can accommodate about a $1\sigma$ excess in the signal events. 

There are several other recent analyses over events $pp\rightarrow h_{i}\rightarrow AA \rightarrow bb\mu\mu$ \cite{CMS:2018nsh} and $pp\rightarrow tH^{\pm}$ followed by $H^{\pm}\rightarrow AW^{\pm}$ and $A\rightarrow \mu\mu$ \cite{CMS:2019idx}. These analyses can probe LS-THDM for relatively small $\tan\beta$ region ($\tan\beta \lesssim 20$) due to its suppression on the coupling $Y_{d}^{u}$. The latter yields a bound on the decay channels involved in the events such that $B_{sig}\equiv {\rm BR}(t\rightarrow bH^{\pm})\times {\rm BR}(H^{\pm}\rightarrow AW^{\pm})\times {\rm BR}(A\rightarrow \mu\mu) \simeq 8\times 10^{-6}$.

\begin{table}[t!]
\centering
\setstretch{1.5}
\scalebox{0.8}{
\begin{tabular}{|l|l|l|l|l|l|l|}
\hline  & Point 1 & Point 2 & Point 3 & Point 4 & Point 5 & Point 6 \\ \hline
$\lambda_{1}$, $\lambda_{2}$ & 0.176, 0.135 & 0.174, 0.13 & {\color{red}-0.337}, 0.111 & {\color{red}-0.141}, 0.387 & {\color{red}-0.036}, 0.131 & {\color{red}-0.409}, 0.031 \\
$\lambda_{3}$, $\lambda_{5}$ & 0.328, 0.327 & 0.394, 0.415 & -0.268, 0.23 & 0.374, 0.243 & {\color{red}0.264}, 0.574 & {\color{red}-0.005}, 0.244 \\
$m_{3}^{2}$ & -1775 & -2126 & -890.1 & -672.2 & -973 & -338.6 \\
$\tan\beta$ & 11.6 & 12.2 & 16.9 & 23.3 & 37.1 & 47 \\ \hline
$m_{1}^{2}$ & 0.327 & 0.177 & -0.451 & -1.5 & 0.02 & -0.003 \\
$m_{2}^{2}$ & -8217 & -7913 & -6728 & -23617 & -8010 & -1919 \\
$\alpha_{H}$ & 0.1$\pi$ & 0.05$\pi$ & -0.1$\pi$ & -0.1$\pi$ & 0.02$\pi$ & {\color{red}0} \\  \hline
$m_{h_{1}}$ & 126.1 & 124.7 & {\color{red}113.2} & 125.3 & {\color{red}126.3} & {\color{red}60.3} \\
$m_{h_{2}}$ & 146 & 162.6 & {\color{red}124.7} & 217.5 & {\color{red}189.5} & {\color{red}126.2} \\
$m_{A}$ & {\color{red}28.1} & {\color{red}28.1} & {\color{red}28.4} & {\color{red}28.6} & {\color{red}29} & {\color{red}28.3} \\
$m_{H^{\pm}}$ & {\color{red}104.3} & {\color{red}116.6} & 90.3 & 91.6 & 136.3 & 92.3 \\ \hline
${\rm BR}(B_{s}\rightarrow \mu^{+}\mu^{-})$ & $ 3.44 \times 10^{-9} $ & $ 3.43 \times 10^{-9} $ & $ 3.55 \times 10^{-9} $ & $ 3.66 \times 10^{-9} $ & $ 3.7 \times 10^{-9} $ & $ 4.15 \times 10^{-9} $ \\
${\rm BR}(B\rightarrow X_{s}\gamma)$ & $3.14\times10^{-4}$ & $3.14\times10^{-4}$ & $3.14\times10^{-4}$ & $3.14\times10^{-4}$ & $3.15\times10^{-4}$ & $3.14\times10^{-4}$ \\
$\xi^{2}{\rm BR}(h_{1}\rightarrow bb)$ & 0.173 & 0.292 & 0.001 & 0.0 & 0.005 & 0.006 \\
$\xi^{2}{\rm BR}(h_{1}\rightarrow \tau\tau)$ & 0.31 & 0.162 & 0.01 & 0 & 0 & 0.0 \\
$B_{{\rm sig}}$ & 0.0 & {\color{red}$ 2.62 \times 10^{-6} $} & 0.0 & 0.0 & {\color{red}$ 1.05 \times 10^{-7}$} & 0.0 \\
$\sigma(pp\rightarrow h_{1}\rightarrow AA \rightarrow bb\mu\mu)$ & 0.003 & 0.018 & 0.021
& 0.0 & 0.0 & 0.0 \\
$\sigma(pp\rightarrow h_{2}\rightarrow AA \rightarrow bb\mu\mu)$ & 0.003 & 0.013 & 0.011
& 0.0 & 0.0 & 0.0 \\ \hline
$\sigma(pp\rightarrow bbA)$ (8 TeV) & 0.125 & 0.112 & 0.059 & 0.031 & 0.012 & 0.008 \\
$\sigma(pp\rightarrow bbA)$ (14 TeV) & 0.243 & 0.218 & 0.114 & 0.06 & 0.024 & 0.015 \\
$\sigma(pp\rightarrow bbA\rightarrow bb\mu\mu)$ (8 TeV) & 0.434 & 0.39 & 0.204 & 0.107 & 0.042 & 0.026 \\
$\sigma(pp\rightarrow bbA\rightarrow bb\mu\mu)$ (14 TeV) & 0.845 & 0.759 & 0.397 & 0.209 & 0.083 & 0.051 \\
S.D. (8 TeV) & {\color{red}1.7$\sigma$} & {\color{red}1.5$\sigma$} & 0.774$\sigma$ & 0.407$\sigma$ & 0.161$\sigma$ & 0.1$\sigma$ \\
S.D. (14 TeV) & {\color{red}2.2$\sigma$} & {\color{red}1.9$\sigma$} & 1$\sigma$ & 0.535$\sigma$ & 0.212$\sigma$ & 0.132$\sigma$ \\ \hline
\end{tabular}}
\caption{Benchmark points exemplifying our findings. All points are chosen to be consistent with the constraints listed in Eq.(\ref{eq:constraints}) and results from the collider analyses. All masses are given in GeV, while the cross-sections are expressed in fb.}
\label{tab:bench}
\end{table}

Before concluding we also present six benchmark points exemplifying our findings in Table \ref{tab:bench}. The standard deviations predicted by the points are shown in the last two rows abbreviated as S.D. supplemented with the COM energy in the parentheses. Points 1 and 2 display solutions which lead to the possible largest excess in the signal events. Point 1 predicts a $1.7\sigma$ excess in 8 TeV COM-collisions, while a $2.2\sigma$ excess for the collisions with 13 TeV COM. $2.2\sigma$ deviation yields slightly more events than observed in \cite{Sirunyan:2018wim}, but considering the uncertainties in our calculations it might still be considered in the acceptable range. Point 2 yields a S.D. at about $1.5\sigma$ in 8 TeV COM-collisions and $1.9\sigma$ for 13 TeV COM-collisions. Point 2 is also subjected to the bound obtained from events involving a light charged Higgs boson \cite{CMS:2019idx}. Points 3, 4 and 5 are compatible with LFU constraints within $1\sigma$; however, the solutions with $\tan\beta \gtrsim 40 $ and $m_{A}\simeq 28$ GeV, as exemplified in Point 6, can be compatible with LFU only within $2\sigma$.  These points are represented rather to reveal the correlation between the observables about the signal and $\tan\beta$. As discussed above, the cross-section $\sigma(pp\rightarrow bbA\rightarrow bb\mu\mu)$ seems to decrease quadratically with increasing $\tan\beta$, but it is still possible to realize the S.D. at about $1\sigma$ for $\tan\beta \sim 17$ as exemplified with Point 3. The S.D. and the cross-section values are predicted less than about $0.5\sigma$ for larger $\tan\beta$ as shown with Points 4, 5 and 6. The solutions with larger $\tan\beta$ need further improvements other than COM and luminosity, since the signal strength is dramatically low due to the $\tan\beta$ suppression. On the other hand, the detector sensitivity to muons have been improved significantly \cite{ATLAS:2019jvq,ATLAS:2020auj}, and the analyses can provide some results even if the signals yield only a few events. In this context, High Luminosity LHC (HL-LHC) can extend the $\tan\beta$ range to about 20 in future analyses.

Since the spectra obtained in LS-THDM is almost the same as the SM except the light Higgs bosons, the solutions exemplified with the benchmark points can be considered to be possible candidates that accommodate the observed excess in the invariant mass of the muon pair. Note that even though we do not consider the alignment limit directly, the LEP bound employed in our analyses constrain the couplings of the non-standard Higgs bosons to the gauge bosons, bottom quark and $\tau-$lepton, which lead to a small $\alpha_{H}$ \cite{Asner:2013psa}. Indeed, the exact limit for the alignment requires $\alpha_{H}$ to be zero \cite{Carena:2014nza,Hou:2017hiw}, which is satisfied by Point 6. The other benchmark points represented in Table 2 satisfy the alignment limit approximately. The electroweak precision data \cite{Baak:2012kk} bound $\alpha_{H}$ as $\sin(\alpha_{H}) \lesssim 0.3$. The largest $\sin(\alpha_{H})$ value happens for the solutions exemplified with Points 1, 3 and 4 which are at the edge of the limit brought by the electroweak precision data.

\section{Conclusion}
\label{sec:conc}

We consider the Higgs boson mass spectrum in the LS-THDM framework and discuss possible impacts from recent collider analyses. We realize that both the CP-odd and CP-even Higgs boson can be light; however, such solutions with $m_{h_{1}}\lesssim 55$ GeV are excluded by the LEP bound obtained from $e^{+}e^{-}$ collisions when $m_{A}\sim 28$ GeV. The spectra involving light Higgs bosons receive an impact from several processes in which the decay modes of the CP-odd Higgs boson into a pair of leptons is involved. The analyses have rather resulted in no significant excess; thus, even though they can constrain the LS-THDM in the low $\tan\beta$ region, they do not provide any process that can probe light Higgs bosons. On the other hand, if one considers the CP-odd Higgs boson production in association with $b-$quarks, an excess observed by the CMS collaboration at about $m_{\mu\mu}\sim 28$ GeV can be applied to the solutions to probe the light Higgs boson mass. The excess has been reported as large as $4.2\sigma$ in the analyses over the collision data with 8 TeV COM, while it is reduced to about $2\sigma$ when 13 TeV COM-collisions are considered. Before probing the solutions with $m_{A}\sim 28$ GeV with the observed excess, one needs to be aware that the same analyses provide also an upper bound for the events involving the $A\rightarrow \tau\tau$ decay channel. The bound restricts the cross-section as $\sigma(pp\rightarrow bbA \rightarrow bb\tau\tau)\lesssim 140$ fb, and it can exclude a significant portion of the parameter space in the same region in which an excess in $m_{\mu\mu}$ can be observed. After constraining the light $m_{A}$ region with consistent ranges of $\sigma(pp\rightarrow bbA \rightarrow bb\tau\tau)$, a largest excess at about $1.5\sigma$ in 8 TeV COM and $2\sigma$ in 13 TeV COM is observed for $\tan\beta \sim 12$ and $m_{A}\sim 28$ GeV. The excess decreases with increasing $\tan\beta$, but it is still possible to realize an excess at about $1\sigma$ for $\tan\beta \sim 17$. We also show that the current sensitivity of the collisions is not enough to probe the large $\tan\beta$ region in LS-THDM; however, this region can be considered in the collisions with higher COM energy and/or more sensitive analyses. Finally, we conclude our discussion by presenting six benchmark points exemplifying our findings.
\acknowledgments
CSU thanks the Physics and Astronomy Department and Bartol Research Institute of the University of Delaware where part of this work has been done. The research of C.S.U. was supported in part by the Spanish MICINN, under grant PID2019-107844GB-C22. Part of the calculations reported in this paper were performed at the National Academic Network and Information Center (ULAKBIM) of TUBITAK, High Performance and Grid Computing Center (TRUBA Resources).

\bibliographystyle{JHEP}
\bibliography{THDMLS}

\end{document}